# Advancing Large Language Models for Spatiotemporal and Semantic Association Mining of Similar Environmental Events

Yuanyuan Tian[1], Wenwen Li[1], Lei Hu[1], Xiao Chen[1], Michael Brook[2], Michael Brubaker[2], Fan Zhang[3], Anna K. Liljedahl[4]

[1] School of Geographical Sciences and Urban Planning, Arizona State University, Tempe, AZ, USA.
[2] Alaska Native Tribal Health Consortium, Alaska, AK, USA.
[3] Bentley University, Boston, MA, USA.
[4] Woodwell Climate Research Center, MA, USA.

**Correspondence**
Wenwen Li
Email: wenwen@asu.edu

# Abstract

Retrieval and recommendation are two essential tasks in modern search tools. This paper introduces a novel retrieval-reranking framework leveraging Large Language Models (LLMs) to enhance the spatiotemporal and semantic associated mining and recommendation of relevant unusual climate and environmental events described in news articles and web posts. This framework uses advanced natural language processing techniques to address the limitations of traditional manual curation methods in terms of high labor cost and lack of scalability. Specifically, we explore an optimized solution to employ cutting-edge embedding models for semantically analyzing spatiotemporal events (news) and propose a Geo-Time Re-ranking (GT-R) strategy that integrates multi-faceted criteria including spatial proximity, temporal association, semantic similarity, and category-instructed similarity to rank and identify similar spatiotemporal events. We apply the proposed framework to a dataset of four thousand Local Environmental Observer (LEO) Network events, achieving top performance in recommending similar events among multiple cutting-edge dense retrieval models. The search and recommendation pipeline can be applied to a wide range of similar data search tasks dealing with geospatial and temporal data. We hope that by linking relevant events, we can better aid the general public to gain an enhanced understanding of climate change and its impact on different communities.

**Keywords**
Climate change; Recommender system; Information retrieval; Semantic similarity; GeoAI; LLM; ChatGPT

# 1. Introduction

The impacts of climate change unfold through a vast number of environmental events that carry profound implications for ecosystems and human societies worldwide. These events cover a broad spectrum, from extreme weather phenomena to gradual ecological succession (Ebi et al., 2021), affecting various regions across different timescales (Hase et al., 2021). Investigating similar environmental events is essential for sharing observations, raising awareness, precepting risks, estimating frequency of unusual events, learning lessons from past occurrences, as well as developing response strategy to climate (Albright & Crow, 2019; Huber & Gulledge, 2011; Raymond et al., 2020).

Despite traditional structured data from sensors, unstructured data such as natural language text can provide valuable insights that structured data alone cannot capture, thus offering a more comprehensive understanding of environmental events (Demanega et al., 2021; Fan et al., 2020; Feng et al., 2022). Mining events concerning climate change typically uses unstructured data from authorized sources such as news articles (Olteanu et al., 2015), or in the form of Volunteered Geographic Information (VGI) (Elwood et al., 2012; Goodchild, 2007) including social media and community reports (Kankanamge et al., 2019; Mihunov et al., 2020; Müller-Hansen et al., 2023; S. Zhang et al., 2021) which contains rich semantics that refers to local knowledge, individual experiences, and local communities impacted by climate change.

Manually curating environmental events and identifying similar events is practical with help from domain experts (Ramachandran et al., 2016). However, it suffers from low efficiency, subjective biases, scalability issues, and lack of scientific rigor and comprehensiveness (Olteanu et al., 2019). In the era of big data, the volume of information on climate change is vast and continuously expanding, and reliance on manual methods is impractical and inadequate. To address these challenges, there's a growing recognition of the merit of utilizing Natural Language Processing (NLP) to automate the process of harnessing and analyzing environmental events (Effrosynidis et al., 2022). As a pivotal subset of artificial intelligence (AI), NLP can efficiently process large textual datasets and plays an important role in advancing Geospatial Artificial Intelligence (GeoAI) (W. Li, 2020) research such as spatial data search and retrieval (Janowicz et al., 2020; W. Li et al., 2011).

Earlier works on geospatial semantic search rely on NLP techniques to compute thematic content similarity and expand geospatial terms with gazetteers. For instance, semantic-enabled indexing and query understanding enable more effective spatial data discovery (W. Li et al., 2014, 2019). Another example is metadata topic harmonization and semantic search driven by linked data for GIS portals (Hu et al., 2015), increasing the accessibility of shared geospatial resources including data and services. NLP can also help with keyword refinement, metadata augmentation, and summarization for geographical data understanding and retrieval (Ferrari et al., 2024; Hu et al., 2019; Tian et al., 2023). These methods laid the groundwork for understanding spatial relationships and thematic relevance within large datasets, forming a crucial foundation that recent LLM applications continue to build upon.

Recently, the emergence of Large Language Models (LLMs) opens a new era of NLP technology with impressive performance on a wide range of natural language tasks. Some recent works have demonstrated the potential of LLMs to assist climate change event analysis such as geographical entity extraction from disaster narratives (Hu et al., 2023) and climate question answering (Thulke et al., 2024; Vaghefi et al., 2023), yet the topic of recommending similar (environmental) events by analyzing spatiotemporal data has not been well studied. The transition from traditional NLP methods to LLMs shows great potential in further enhancing geospatial semantic search and recommender systems.

Leveraging NLP for identifying similar environmental events is a recommendation problem in the realm of information retrieval (IR). The primary objective is to fetch relevant items from a large set of candidates for the query item (Lu et al., 2015). Traditional IR methods have relied on keyword-matching techniques such as BM25 algorithm (Robertson & Zaragoza, 2009), which is simple and efficient based on item description, but falls short in capturing the deeper semantic relationships between words and phrases. Semantic search (Bast et al., 2016) is a more advanced approach that goes beyond mere keyword-based search methods by paying attention to contexts, thus enabling a more refined retrieval of information. Semantic search has been significantly advanced by recent development and application of LLMs such as Bidirectional Encoder Representations from Transformers (BERT) (Devlin et al., 2018), Generative Pre-trained Transformer (GPT) models (Brown et al., 2020), and Pathways Language Model (PaLM) (Chowdhery et al., 2022).

Although semantic search has significantly enhanced IR capabilities, solely relying on it risks neglecting critical spatial and temporal dimensions of environmental events. A substantial portion of unstructured data, particularly from news and social media, is inherently enriched with geospatial and temporal information and metadata. For environmental events, their geographical location and time of occurrence are important facets enabling spatiotemporal analysis. Disregarding these aspects can lead to a fragmented understanding of environmental phenomena, hindering efforts to accurately identify, link, and analyze similar events. This limitation indicates a critical gap in current NLP solutions, emphasizing the need for a framework for event recommendation that extends beyond mere semantic search to encompass spatiotemporal dynamics (Buckingham et al., 2020).

To overcome the limitations of current methods, we propose a framework that integrates LLMs with spatiotemporal proximities as well as other semantic and spatial relevance measures to enhance the process of identifying similar spatiotemporal events. By leveraging advanced NLP techniques, our framework successfully captures the spatial, temporal, and semantic dimensions of spatiotemporal phenomena, providing scalable and automated assistance to manual curation. At the heart of this framework is the Geo-Time Re-ranking (GT-R) model, designed to surface events by evaluating their semantic content, spatial proximity, and temporal relevance. This innovative integration advances geoinformation retrieval and environmental informatics. It can facilitate the analysis of extensive datasets to reveal critical patterns and connections essential for devising effective climate change adaptation and mitigation strategies.

Contributions of this work are summarized below:
(1) We proposed a two-stage LLM-based search and recommendation framework designed for analyzing spatiotemporal events. This framework uniquely incorporates spatial and temporal facets in both retrieval and re-ranking stages.
(2) We introduced an innovative use of LLMs to enhance the category feature of events through zero-shot named entity recognition, enriching the semantics with minimal human intervention.
(3) We proposed and developed the GT-R model, a novel re-ranking solution that synthesizes semantic understanding with geographical and temporal relevance.
(4) We provided empirical evidence demonstrating the superior performance of our proposed framework compared to existing dense retrieval and cutting-edge re-ranking models, highlighting its potential to significantly improve spatiotemporal event recommendation.

The remainder of this paper is organized as follows to provide a comprehensive exploration of our framework and findings. Section 2 reviews related work, covering the utilization of LLMs for recommendation systems, domain-specific adaptations related to climate change, and the integration of spatiotemporal dimensions in news recommendation. Section 3 details our proposed method, which harnesses spatiotemporal insights along with LLMs to enhance the accuracy and relevance of event recommendations. Section 4 outlines the experiments to validate our method and compare it against alternative approaches. Section 5 presents the experiment results, showcasing the effectiveness of our approach in environmental event recommendations. Section 6 discusses the broader implications of our findings, particularly the potential of LLMs to transform recommendation systems through a spatiotemporal lens. Finally, Section 7 concludes the paper by summarizing our contributions and suggesting avenues for future research.

# 2. Related work

## 2.1 Large language models and recommendation

Recommendation, as a fundamental task in information retrieval, demands a thorough understanding of content relevance to tackle challenges such as the cold-start problem (Arora et al., 2023), where user interactions with content are limited or nonexistent. This issue highlights the importance of a well-founded recommendation system architecture. A common approach in recommendation system design employs a multi-stage ranking pipeline that typically starts with retrieval followed by re-ranking (Yue et al., 2023). In the evolving landscape of NLP and AI research, LLMs have gained significant attention for their strong capabilities across various tasks (Zhu et al., 2023). Recently, researchers have started to investigate how LLMs can advance these stages of recommendation in various ways (Lin et al., 2023; Wang et al., 2023; L. Wu et al., 2023).

The retrieval phase aims to shortlist candidates from extensive datasets, typically through sparse retrieval techniques, or dense retrieval utilizing LLM embeddings for semantic search (Gao et al., 2023). LLMs are particularly good at capturing data semantics by generating dense representation via bi-encoding (Dai et al., 2022), due to their deep semantic understanding and contextual awareness, thereby improving the relevant item retrieval process. The subsequent re-ranking phase is refined by leveraging LLMs for their ability to analyze complex relationships. This is often achieved through the cross-encoding architecture (Askari et al., 2023; Reimers & Gurevych, 2019), facilitating pointwise similarity calculation, i.e., encoding the query and candidate results simultaneously. MonoRank was the first work that uses cross-encoder based on BERT model for re-ranking (Nogueira & Cho, 2019). Following that, other works have explored more effective ways to conduct re-ranking through fine-tuning LLMs using the Microsoft MARCO dataset (Bajaj et al., 2016). Recently, the information retrieval community has explored the possibility of directly employing LLMs for re-ranking based on token (e.g., word) prediction through various methods including pointwise, pairwise, and listwise approaches (Qin et al., 2023; Sun et al., 2023; Zhuang et al., 2023), each varying in the comparison scope of candidates.

Despite LLMs' strong language processing capabilities, the efficacy of retrieval using LLM embedding models can be influenced by the input data structure. For example, providing a proper context by adding or tuning prefix improves LLM performance without changing its parameters (X. L. Li & Liang, 2021). A recent code-driven reasoning research shows that LLMs can gain performance improvement by reformatting contents in a syntactic structure (C. Li et al., 2023). Moreover, most existing LLM-based re-ranking efforts focus predominantly on the semantic aspects, paying little attention to other critical contextual dimensions such as geographical location and temporal factors. In our work, we explore the optimal way to formalize contents to make the most effective use of LLM embedding models during the retrieval stage, and the potential application of LLMs in the re-ranking stage. Combining both strategies, our goal is to improve the effectiveness of spatiotemporal event recommendation.

## 2.2 Domain adaptation and climate change research

Domain adaptation of LLMs is crucial for tailoring these general models to a specific domain without the need for training models from scratch. Prompting emerges as a pragmatic and accessible method for achieving domain adaptation (Rao et al., 2023), employing suitable prompts and instructions to gear LLMs toward generating outputs aligned with the desired context or task without altering LLMs' underlying parameters. For example, prompt engineering based on domain expertise can improve the performance of LLMs because it leverages the model's existing knowledge, providing context, instructions, and examples at inference time to produce relevant responses (Vaghefi et al., 2023). The approach varies from zero-shot learning, requiring no specific examples, to few-shot learning which provides a few examples (Brown et al., 2020; Wei et al., 2021). Additional strategies to enhance prompt engineering include persona adoption (Thulke et al., 2024), which assigns the model with a predefined character or perspective, and the Chain of Thought (CoT) method, which guides the model through a step-by-step reasoning process (Wei et al., 2022). Alternatively, fine-tuning adjusts the model's parameters via training on a domain-specific dataset to help with tasks such as climate-related text classification and geography-related toponym recognition (Webersinke et al., 2021; Zhou et al., 2023). Both methods leverage the intrinsic capabilities of LLMs and enable LLMs to potentially serve the complexities of climate change research.

Zero-shot learning has shown remarkable efficacy for domain adaptation in climate change research. Focusing more on high-quality instructions rather than searching for the extensive and unbiased example data required by continuous training or fine-tuning methods, it represents a more intuitive and resource-efficient approach (Zhong et al., 2021). For instance, using zero-shot learning, recent works have shown that LLMs can even outperform human annotators on news or social media posts, or more specifically, help with classification of climate-related disclosures, detecting public options regarding climate activism, with minimal domain-specific training (Auzepy et al., 2023; Emran et al., 2024; Gilardi et al., 2023). This capability has been pivotal in enhancing the accuracy and contextual relevance of information retrieval and analysis in environmental research and climate change studies, showcasing the transformative potential of adapting LLMs to specific fields. Based on this review, our work employs various prompt engineering techniques to utilize LLMs in building new features to assist with re-ranking.

## 2.3 News recommendation and spatiotemporal dimensions

Event recommendation, particularly within the realm of news recommendation, emerges as a multifaceted problem. Content serves as the primary foundation, and content features can be constructed in terms of semantic, entity, keyword, topic, emotion, and other factors (Jiang et al., 2021; C. Wu et al., 2023). Recent advancements in LLMs have enhanced the capabilities of news recommendation systems, offering new methods for feature construction and contextual understanding (Q. Liu et al., 2023, 2024). These advancements allow for the efficient integration of various contextual aspects, demonstrating the substantial benefits of incorporating not only the content itself but also the context in which news events occur. Such contextually enriched approaches have led to more relevant news recommendations even in the cold start setting (Q. Liu et al., 2024), leveraging LLMs' advanced NLP capabilities to parse and understand complex datasets.

The incorporation of spatiotemporal perspectives into news recommendation systems introduces an additional dimension of relevance by employing location and time as critical factors. While attempts to integrate spatiotemporal factors have utilized geotags and timestamps, most applications focus on depicting user context and behavior (Javed et al., 2021; Mitova et al., 2023; Raza & Ding, 2022). Location and date of news have often been overly simply transferred as property features such as distance for radius search, start and end date for time range filtering or recency (Bao & Mokbel, 2013; Shah et al., 2023; Xu et al., 2015). The potential for deeper domain-specific insights into spatial and temporal similarities remains overlooked. For instance, environmental events are inherently linked to their geographical location and time of occurrence. In that sense, they can exhibit latitudinal similarities (Malizia et al., 2020; Nishizawa et al., 2022). Their occurrences might be seasonal rather than strictly date-specific. Examples are yearly wildfires in California and temperature-correlated peak thawing of permafrost in Alaska. Although LLMs have demonstrated efficiency as semantic re-ranker (Sun et al., 2023), using LLMs as a sole solution is limited in handling the complexities of space and time dimensions adequately. Moreover, recent studies show that LLMs possess an inherent understanding of space and time in the real world and the capability to consume location and timestamp (Gurnee & Tegmark, 2023; Roberts et al., 2023; Z. Zhang et al., 2024), showing the potential to process content with geographical and chronological references.

Although there are GeoAI research trying to incorporate spatial or temporal thinking into large deep learning models (W. Li et al., 2021; Luo et al., 2023; Xie et al., 2023), very little relevant research has been done in LLMs area, particularly in news recommendation (Hu et al., 2023). To address these challenges, our proposed framework explores news recommendations within the context of climate change related events. This new framework incorporates multi-faceted criteria, including the critical spatial, temporal, and categorical, in addition to the popular semantic similarity measures to better understand, and perform search and recommendation of relevant events/news related to climate change.

# 3. Method

In this section, we first introduce the proposed framework for the search problem regarding retrieving relevant events with spatiotemporal information. Then, we dive into details of constructing each feature including semantic similarity calculation, category similarity calculation, relevance calculation in terms of space and time. Finally, we explain how to combine those feature lists to form the final ranking list of candidate events.

## 3.1 Overall framework

In this section, we introduce a novel framework designed to tackle the challenge of mining relevance of spatiotemporal events/news by pinpointing events that are similar not only in content but also in their spatiotemporal context. As depicted in Figure 1, our framework is structured into two interconnected stages. Initially, the framework requires structuring the event descriptions. This tailors the content to more closely reflect user intent by strategically integrating prefixes and enhancing the descriptions with spatiotemporal metadata, thereby enriching the context. The first stage is dedicated to retrieval. Leveraging an LLM embedding model, the system retrieves a collection of events that are preliminarily relevant to the query based on their semantics. Subsequently, in the second stage, we introduce a refinement process as re-ranking. This stage uses the proposed GT-R model, which unravels the intricacies of semantic pertinence and spatiotemporal coherence. The GT-R model re-evaluates the preliminary set, assessing each candidate event's relevance through a fusion of scores derived from semantic and spatiotemporal features. The outcome of this process is the generation of a final, re-ranked set of events. This set represents the most contextually and temporally relevant events and is presented as the ultimate output of our recommendation framework. This carefully orchestrated sequence ensures that the retrieval system identifies events with relevant information with the combined spatiotemporal search criteria.

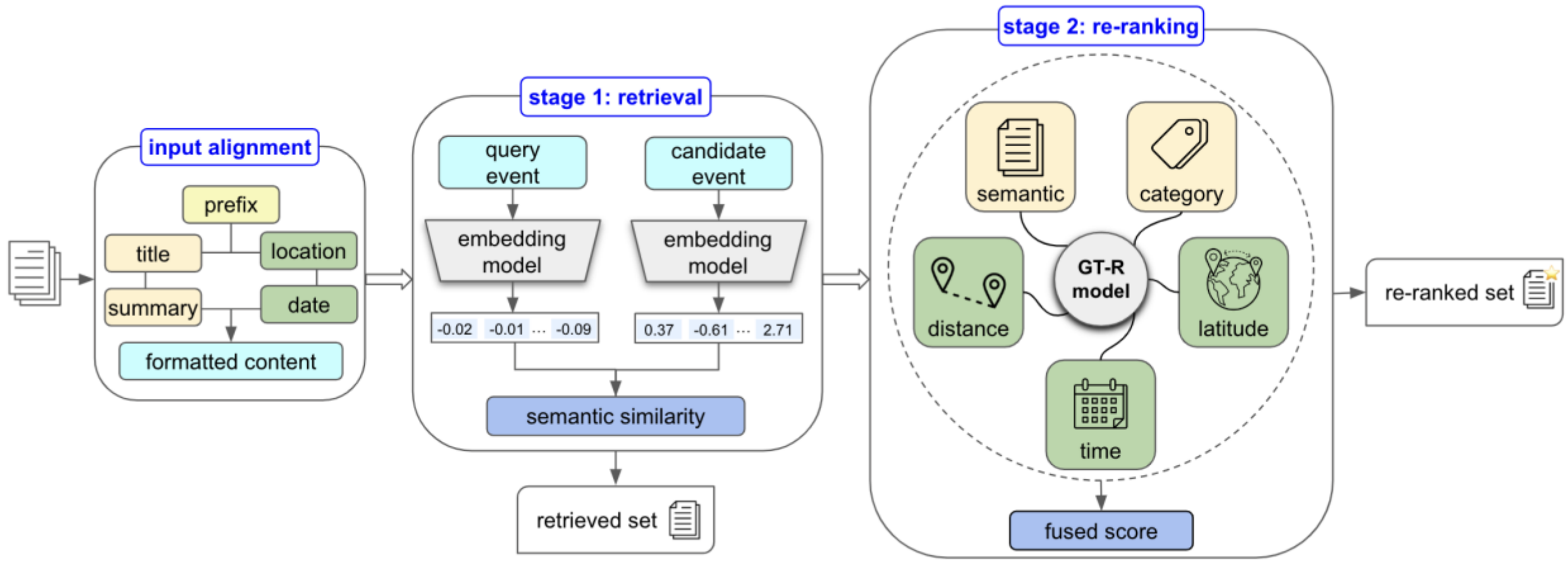

**Figure 1.** The proposed framework for mining relevant spatiotemporal news events.

## 3.2 Feature construction

We detail the methodologies for calculating semantic similarity, category-instructed similarity, spatiotemporal relevance, and the subsequent fusion of these features to construct the final ranking of related events. Algorithm 1 describes the formulation of the GT-R model and its annotations.

---

**Algorithm 1.** GT-R: Geo-Time Re-Ranking

---

**Input**:

Query $q$, corpus $Z$ (the cardinality of $Z$ is $N_z$),
number of candidates to retrieve $N_{\text{retrieve}}$, number of candidates to re-rank $N_{\text{re-rank}}$,
bi-encoding similarity model $f_{\text{sim_bi}}$, cross-encoding similarity model $f_{\text{sim_cross}}$,
category-instructed entity extraction model $f_{\text{NER}}$, distance calculation model $f_{\text{distance}}$,
latitude calculation model $f_{\text{latitude}}$, time calculation model $f_{\text{time}}$,
semantic similarity weight $\tau_s$, category similarity weight $\tau_c$,
distance threshold $\tau_d$, distance boosting factor $\beta_d$, latitude threshold $\tau_\phi$, latitude boosting factor $\beta_\phi$.

**Initialize**:

retrieved event set $Z_{\text{retrieve}} \leftarrow \{\}$, output re-ranked event set $Z_{\text{re-rank}} \leftarrow \{\}$.
Temporary set $Sim_s \leftarrow \{\}$, Temporary set $Sim_c \leftarrow \{\}$,
Temporary set $D \leftarrow \{\ \}$, Temporary set $L \leftarrow \{\ \}$, Temporary set $T \leftarrow \{\ \}$. Temporary set $RRF \leftarrow \{\ \}$.

**do**

Calculate semantic similarity scores: $\forall z \in Z, Sim_s \leftarrow Sim_s \cup \{f_{\text{sim_bi}}(q, z)\}$.

The cardinality of $Sim_s$ is $N_z$. Denote $Sim_s = \{s_1, s_2, \dots s_{N_z}\}$.

Sort (descending) and keep top $N_{\text{retrieve}}$ candidates: $Z_{\text{retrieve}} \leftarrow \{z \mid f(q, z) = s_{(i)}, i \leq N_{\text{retrieve}}\}$ and $s_{(i)}$ is the $i$-the largest similarity score of the sample.

① Semantic similarity

Obtain the raw ranking order of semantic similarity scores: $R^0_{\text{semantic}} \leftarrow \{r^d_{\text{semantic}}(s): s \in Sim_s\}$,
where $r^d_{\text{semantic}}(\cdot)$ gets the descending rank of semantic similarity score s.

Apply weight to adjust $R_{\text{semantic}} \leftarrow \{\hat{r}_{\text{semantic}}(z) = \frac{r}{\beta_s} \mid r \in R^0_{\text{semantic}}\}$.

② Category similarity

Calculate category similarity scores: $\forall z \in Z_{\text{retrieve}}, Sim_c \leftarrow Sim_c \cup \{f_{\text{sim_cross}}(f_{\text{NER}}(q, z))\}$.

Obtain the ranking of category similarity scores: $R^0_{\text{category}} = \{r^d_{\text{category}}(s): s \in Sim_c\}$,
where $r^d_{\text{category}}(\cdot)$ gets the descending rank of category similarity scores $s$.

---

Apply weight to adjust $R_{\text{category}} \leftarrow \{\hat{r}_{\text{category}}(z) = \frac{r}{\beta_c} \mid r \in R^0_{\text{category}}\}$.

③ **Distance relevance**

Calculate the distances: $\forall z \in Z_{\text{retrieve}}, D \leftarrow D \cup \{f_{\text{distance}}(q, z)\}$.

Obtain the ranking of distances: $R^0_D = \{r^a_{\text{distance}}(s): s \in D\}$,
where $r^a_{\text{distance}}(\cdot)$ gets the ascending rank of the distance $s$.

Adjust $R_{\text{distance}} \leftarrow \{\hat{r}_{\text{distance}}(z) \mid r \in R^0_D\}$, where

$$\hat{r}_{\text{distance}}(z) = \begin{cases} r / \beta_d, & \textbf{if } f_{\text{distance}}(q, z) < \tau_d \\ r, & \textbf{if } f_{\text{distance}}(q, z) \geq \tau_d \end{cases}$$

④ **Latitude relevance**

Calculate the latitude differences: $\forall z \in Z_{\text{retrieve}}, L \leftarrow L \cup \{f_{\text{latitude}}(q, z)\}$.

Initialize the ranking of latitude using raw ranking: $R^0_L \leftarrow R^0_{\text{semantic}}$.

Adjust $R_{\text{latitude}} \leftarrow \{\hat{r}_{\text{latitude}}(z) \mid r \in R^0_L\}$, where

$$\hat{r}_{\text{latitude}}(z) = \begin{cases} r / \beta_\phi, & \textbf{if } f_{\text{distance}}(q, z) \geq \tau_d \textbf{ AND } |f_{\text{latitude}}(q, z)| < \tau_\phi \\ r, & otherwise \end{cases}$$

⑤ **Temporal relevance**

Calculate the temporal relevance: $\forall z \in Z_{\text{retrieve}}, T \leftarrow T \cup \{f_{\text{time}}(q, z)\}$.

Obtain the ranking of temporal relevance: $R^0_T = \{r^a_{\text{temporal}}(s): s \in T\}$,
where $r^a_{\text{temporal}}(\cdot)$ gets the ascending rank of the temporal $s$.

Get ranking: $R_{\text{time}} \leftarrow \{\hat{r}_{\text{temporal}}(z) = r^a_{\text{temporal}}(s)\}$.

**Fuse rankings** ①②③④⑤

Form ranking set $R \leftarrow \{R_{\text{semantic}}, R_{\text{category}}, R_{\text{distance}}, R_{\text{latitude}}, R_{\text{time}}\}$.

Obtain fused score. $\forall z \in Z_{\text{retrieve}}, RRF \leftarrow RRF \cup \{f_{RRF}(z)\}$, where $f_{RRF}(z) = \sum_{r \in R} \frac{1}{k + r(z)}$.

Obtain the ranking for the fused score: $R_{RRF} = \{r^d(s): s \in RRF\}$,
where $r^d(\cdot)$ gets the descending rank of the RRF score $s$.

$Z_{\text{re-rank}} \leftarrow \{z \in Z_{\text{retrieve}} \mid r^d(f_{RRF}(z)) \leq N_{\text{re-rank}}\}$.

**return** $Z_{\text{re-rank}}$.

### 3.2.1 Semantic similarity

The fundamental goal is to achieve an enhanced understanding of the semantics underlying event descriptions. Basic methods such as BM25 are good at matching keywords but often fail to capture the context necessary for discerning relevance among complex events. This calls for a more sophisticated model that can address overt and subtle contextual signals within the data, which can potentially improve retrieval performance. Following advances in NLP, particularly the success of transformer-based models (W. Li & Hsu, 2022) in capturing deep semantic meanings, we leverage the advanced LLM embedding model - OpenAI's Ada (Neelakantan et al., 2022) that has shown proficiency in understanding context in long ranges. By employing semantic embeddings, our method captures the important hidden relationships within text, enabling a richer and more accurate representation of content semantics. This approach is further refined through input segment picking and structuring to better align the model's understanding of spatiotemporal events. Specifically, different combinations of content segments are compared using basic concatenation, and further compared the best combination with adding prefix accordingly.

After metadata structuring, a query event $q$ with a candidate event $z$ from the set of all event records $Z$ are fed into the LLM embedding model separately, and mapped as two high-dimensional vectors for semantic similarity calculation as noted by $f_{\text{sim_bi}}$. According to semantic similarity scores $Sim_s$ from large to small, the corpus is ranked as an ordered list of events. Then, the top $N_{\text{retrieve}}$ candidates are shortlisted as the retrieved result $Z_{\text{retrieve}}$ that exhibit the highest degree of semantic consistency. For example, a query containing the title "Massachusetts city got nearly 10 inches of rain in 6 hours, flooding homes and eroding dams" would see higher ranks for events such as the one detailing "Surprise flooding hits several Kelowna neighborhoods …" over broader events such as "Extreme rain" because the close semantics of "city" and "neighborhoods" are captured. This semantic similarity measurement provides a foundation for event recommendation.

### 3.2.2 Category similarity

Categories are invaluable semantic information tagged by domain experts towards thematic similarities that can be leveraged to assist in re-ranking, especially with well-defined categorization schema. While the common faceted search method uses category tags as simple filters, it is unable to capture the semantic depth of the categories solely on word matching. On the other hand, important entities within news content carry key messages and facilitate a more direct understanding of the content. A recent work suggests that by explicitly revealing the category of these entities, the model can understand and process the content more accurately (D. Liu et al., 2020). To address these challenges, our methodology proposes a hybrid model that not only leverages semantics in category tagging but also enriches it by entity extraction as illustrated in Figure 2. The intuition is that incorporating categorized entities into the news representation will provide a more condensed semantic profile of each article, which in turn will enhance the relevance and accuracy of recommendations in a cold-start scenario. Through the application of prompt engineering, we harness the processing power of LLM in zero-shot settings, supplemented by persona setting, task description, and CoT processes, to effectively incorporate category tags into instructions. In this way, domain knowledge representing category tags is injected to an LLM. Recognizing the limitations of generative models such as output might be malformatted even after multiple retries, we incorporate a "Human in the Loop" approach, which involves manual oversight to refine and verify the outputs of this category-entity fusing process by LLM.

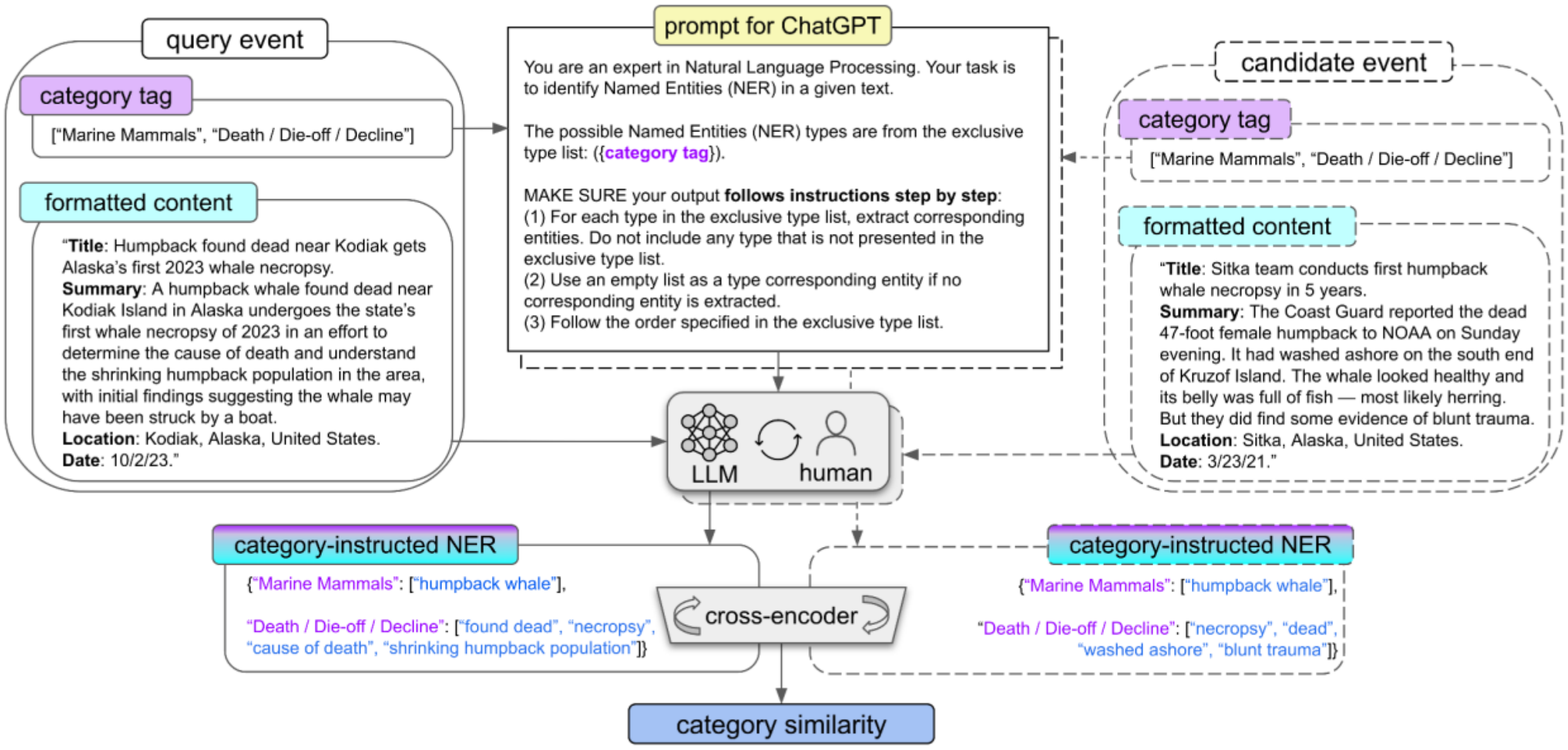

**Figure 2.** An illustration of calculating category similarity in the GT-R model. The example conducts category-instructed entity extraction of a query event and a candidate event respectively using an LLM with human-in-the-loop, followed by cross-encoding for similarity calculation.

To build the category similarity feature, for each pair of a query event $q$ and a candidate event $z$ from $Z_{\text{retrieve}}$, we initiate the process with entity extraction $f_{\text{NER}}$. This extraction is coupled with the human-identified one or more corresponding categories to form two new text snippets. Each snippet integrates category tags with the corresponding entities identified within the event descriptions. In Figure 2, both the query event and the candidate event are tagged with the category "Marine Mammals", and the entity "humpback whale" is extracted from both. This indicates a high degree of topical focus and a fine-grained similarity at the species level. Additionally, with the tag "Death / Die-off / Decline", while one event mentions "blunt trauma" the other does not, signaling a subtle variance within the category. These text snippets are then concurrently processed by the cross-encoder $f_{\text{sim_cross}}$. Different from the bi-encoding architecture used for dense retrieval, which outputs vectors requiring subsequent similarity computations, the cross-encoder directly predicts a similarity score. This method achieves a higher precision than bi-encoding, albeit with increased computational demands so usually applied to limited comparisons. The candidate events in $Z_{\text{retrieve}}$ are then re-ranked based on the descending semantic similarity scores $Sim_c$ . This ranking reflects the depth of category-based semantic connections between events. This dual strategy proposes a more refined and granular approach to establishing category similarity. It allows for an enriched semantic understanding of the core of an event, which helps precise and accurate re-ranking of pertinent events.

### 3.2.3 Spatiotemporal relevance

#### (1) Distance relevance

We consider the first spatial feature based on distance. This comes from Tobler's first law of geography, "Everything is related to everything else, but near things are more related than distant things" (Tobler, 1970), and the notions of weight and scale from Geographically Weighted Regression (Brunsdon et al., 1998). This also applies to biological systems as illustrated by a general and strong occurrence of geographic distance decay in ecological community similarity (Morlon et al., 2008). Our method uses distance as the spatial proximity to measure the closeness of events in a meaningful way than simple radius search. Distance between an event pair is calculated using the Haversine formula which determines the great-circle distance between two points on a sphere given their longitude and latitude coordinates. This is a more accurate measure of distance because some unusual environmental events are so rare that relevant events might be far even at continental scale.

Let $(\phi_q, \lambda_q)$ be the latitude and longitude of where the query event $q$ occurs, and $(\phi_z, \lambda_z)$ be the latitude and longitude for the occuring location of the candidate event $z$. The Haversine distance $D$ between $q$ and $z$ can be calculated using Equation 1. $R$ is the Earth's radius, taken as a constant representing a mean radius of 6,371 km. As shown in Algorithm 1, a distance-based ranking list is built to prioritize events that fall within a certain distance $\tau_d$, coupled with a boosting factor $\beta_d$, thereby enhancing the relevance of search results. This approach ensures an enhanced understanding of spatial relevance is applied to event retrieval, thus fulfilling the objective of providing users with more contextually appropriate search results based on geographical proximity.

$$D(q,z) = 2R \cdot arcsin\left(\sqrt{sin^2\left(\frac{\phi_z - \phi_q}{2}\right) + cos(\phi_q)\, cos(\phi_z)\, sin^2\left(\frac{\lambda_z - \lambda_q}{2}\right)}\right) \tag{1}$$

#### (2) Latitude relevance

Latitude serves as the second geo-related feature, leveraged to capture similarities in climate, weather patterns, species distribution, and behaviors that are often correlated with latitudinal gradients. Studies have demonstrated that species' responses to climate change, including shifts in phenology and potential for extinction, are significantly modulated by their latitudinal position (Deutsch et al., 2008; Francis & Vavrus, 2012; VanDerWal et al., 2013). Furthermore, the generation and maintenance of latitudinal diversity gradients themselves are intricately linked to spatiotemporal climatic variations (Saupe et al., 2019). By incorporating latitude as a key feature, our model aligns with these established ecological and climatological research findings, as well as a recent GeoAI reproducibility research (W. Li et al., 2024) that show similar patterns and conditions are associated with similar latitudinal zones. We introduce a latitude booster to our ranking algorithm, which is conditioned on a distance threshold $\tau_d$. This booster is designed to increase the relevance of events that are not only geographically proximate but also share similar latitudinal characteristics. The boost factor for latitude $\beta_\phi$ is applied to events within a predefined latitudinal range $\tau_\phi$, effectively drawing attention to events that occur within ecologically and climatically similar zones. For instance, an event about an early arrival of Arctic tern in Iceland would be deemed more relevant to an event regarding and early observation of pintail ducks in Alaska than an event with a similar category as “birds” but occurring at a different latitude, such as an event of usual migration routes of whooper swan reached Manitoba. By incorporating latitude into our feature set, we provide a perspective that goes beyond mere geographic coordinates, offering insights into potential environmental and ecological connections between events.

### (3) Temporal relevance

The goal of creating a time-related feature is to refine retrieved events by integrating a temporal dimension that resonates with the occurrence of environmental phenomena. Conventional retrieval systems, which may hinge on specific calendar dates, often overlook the cyclical nature of environmental events. To accommodate this temporal expectation, our methodology adopts the day-of-year as a more meaningful temporal metric, allowing us to rank events in a way that more accurately reflects their environmental significance and timeliness. Events are re-ranked based on this temporal proximity of events noted as the function $f_{\text{time}}$. The temporal ranking reflects the likelihood of events sharing similar environmental conditions or occurrences based on their seasonal timing. It ensures that events occurring within the same seasonal period are recognized as temporally similar. For example, in the region of Cowichan River in British Columbia, Canada, a drought event that happened in May 2019 is more temporally related to another drought event of this river in summer 2017, rather than a drought event recorded in fall, 2018. This feature prioritizes the alignment of events with seasonal cycles, which is particularly applicable for environmental datasets where timing can be critical.

## 3.3 Feature combination and rank fusion

The goal is to optimize the final event rankings, ensuring that the most relevant, contextually and spatiotemporally relevant events are surfaced to the user. Before going into details of the GT-R algorithm, we introduce notations for two statistical concepts: order statistics and rank statistics. Order statistics serve as the backbone for organizing the features derived from spatiotemporal events, allowing us to systematically assess their relevance. By defining a sample $\{X_i\}_{i=1,2,\ldots,n}$ of size $n$, we utilize order statistics to arrange this event list in either descending or ascending order based on the criteria of each feature. The descending order, signified by $X_{(i)}$, denotes the $i$-th largest element, ensuring a hierarchy where $X_{(1)} > X_{(2)} > \cdots > X_{(n)}$. Conversely, the ascending order prioritizes the event from the least to the most significant one, identified as $X_{[i]}$, which represents an inverse order as $X_{[1]} < X_{[2]} < \cdots < X_{[n]}$. Ranking statistics further complements our framework. By applying ranking statistics to a sample $\{Z_i\}_{i=1,2,\ldots,n}$, we can assign a distinct rank based on the event’s significance, either in descending order $R_i^D$, from the most to the least significant, or in ascending order $R_i^A$, from the least to the most significant. Through the function $r^d$ for descending ranks and $r^a$ for ascending ranks, we achieve a permutation of integers that mirrors the complex interplay of events.

After constructing features for re-ranking, disparate features are synthesized into the final ranking list through feature combination and rank fusion. The convex combination approach offers a mathematical framework that allows for the weighted integration of various features (Norouzi et al., 2013), providing a flexible means to emphasize or de-emphasize certain aspects based on their relevance or reliability. Reciprocal Rank Fusion (RRF) is an approach that effectively aggregates rankings from diverse sources (Cormack et al., 2009), thereby reducing the bias and variance associated with individual ranking strategies. Our fusion strategy employs both as the rank fusion methodology that serves GT-R, and its procedure is demonstrated in Algorithm 1. It also adjusts individual features (with or without sorting), selectively applies booster factors, and calibrates the feature weights. This refined fusion strategy is designed to synergistically combine semantic-related features including semantic similarity and category relevance similarity into one variable with adjusted weights, then fuse with spatiotemporal proximity features into a unified ranking. By balancing these components within our rank fusion method, we aim to achieve a harmonized ranking output that captures the multifaceted nature of event relevance in the context of relevant spatiotemporal event retrieval.

# 4. Experiment

This section gives details of experiments regarding dataset, metrics, and comparing models.

## 4.1 Dataset

The experimental dataset is sourced from the Local Environmental Observer (LEO) Network, a collaborative project that shares local news and reports on unusual environmental events, with a focus on the Arctic regions as well as global occurrences (Mosites et al., 2018). Addressing climate change at the human scale, the LEO Network highlights local impacts on communities. The community-driven perspectives (Huntington, 2011) highlighted by the LEO project are essential in effectively tackling climate change challenges (Danielson et al., 2022). The repository of LEO Network events serves to increase public awareness regarding climate dynamics in the Arctic and across the planet.

An example of a LEO event is illustrated in Figure 3 (a). An event is structured and encompasses several content segments: title, summary, date, location, category, and a “see also” section for relevant event recommendations. Event reporting within the LEO Network is predominantly community-driven, with events initially submitted by local observers. These contributions are subsequently refined by experts possessing relevant domain knowledge to ensure the accuracy and quality of the information, which includes the manual curation of related events. This process is labor intensive, so the goal of this use case is harnessing the power of AI to assist automated retrieval of relevant events.

An examination of the lexical elements commonly mentioned in LEO events is visualized in Figure 3(b) through a word cloud. Prominent among these are locational and temporal references (e.g., “Alaska”, “Finland”, “Arctic”, “winter”), meteorological terms (e.g., “storm”, “flooding”, “wildfire”), and species names (e.g., “salmon”, “seal”, “bear”). Event categories are derived from a tagging schema defined by the LEON area experts and consultants based on practical experience (Danielson et al., 2022), ensuring relevance to the environmental domain and climate change. The comprehensive schema includes 100 tags. As demonstrated in Figure 3(c), event category tags ranging from “Natural Environment” tags such as “Ocean/Sea”, to descriptors of “Unusual or Unexpected Events” such as “Extreme Precipitation”, and tags indicating “Impact on Human Environment” such as “Transportation”.

For the purposes of this study, we conducted both quantitative and qualitative evaluations of the proposed GT-R model using the LEO spatiotemporal event dataset. This dataset comprises 1,000 query events, each linked to related events, which were human annotated from a pool of approximately four thousand event candidates from 05/20/1998 to 10/02/2023. Each event comprises content segments of title, summary, location (with longitude and latitude), and date.

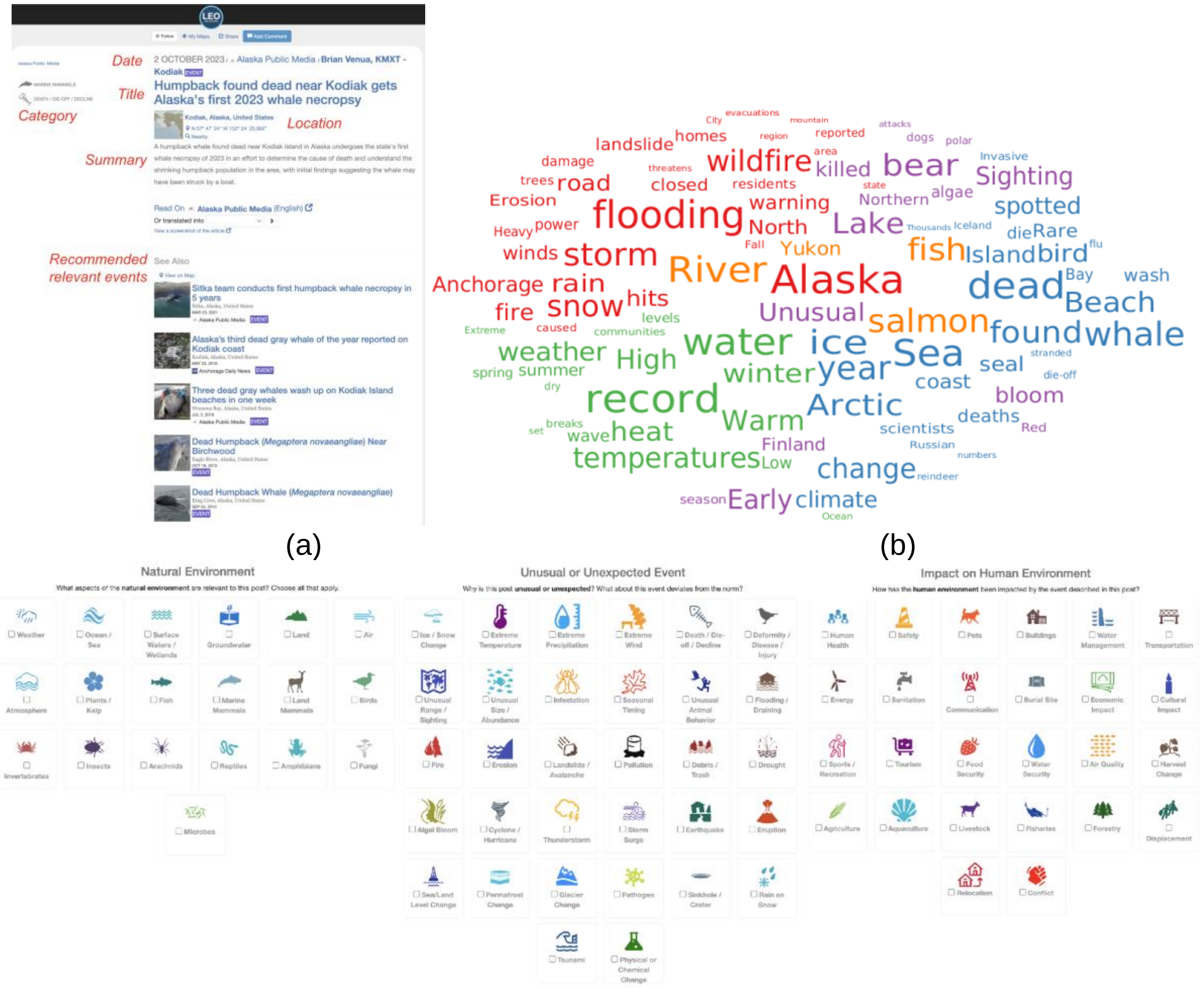

(c)

**Figure 3.** LEO environmental event dataset: (a) an event example with content structure explained; (b) visualization of high-frequency words in event content; (c) event category.
(a) and (c) are screenshots of LEO web-based portal, (b) is generated using event title.

## 4.2 Evaluation metrics

We focused on four standard evaluation metrics in information retrieval tasks, i.e., Recall, Hit Rate, Normalized Discounted Cumulative Gain (nDCG), and Mean Reciprocal Rank (MRR) as calculated using Equation (2-6) to assess model performance:

$$Recall@k = \frac{R_k}{R_{total}} \tag{2}$$

$$Hit\ Rate@k = \frac{H_k}{Q_{total}} \tag{3}$$

$$DCG@k = \sum_{i=1}^{k} \frac{Relevance\ score}{\log_2(i+1)} \tag{4}$$

$$nDCG@k = \frac{DCG@k}{ideal\ DCG@k} \tag{5}$$

$$MRR@k = \frac{1}{k}\sum_{i=1}^{k} \frac{1}{rank_i} \tag{6}$$

In stage 1 (retrieval), which aims to select the top 100 candidates, we employ two key metrics: Recall and Hit Rate. $Recall@k$ measures the fraction of relevant events retrieved within the highest-ranking $k$ positions of the result set. It is formally defined as the ratio of the number of relevant events retrieved in the top $k$ positions ($R_k$) to the total number of relevant events available within the dataset ($R_{total}$) (Equation 2). $Hit\ Rate@k$ reflects the frequency with which the retrieved set includes at least one relevant event within the top $k$ positions for a given query event. It is calculated as the number of query events with at least one relevant event in the top $k$ positions ($H_k$) divided by the total number of query events ($Q_{total}$) (Equation 3). Both metrics are crucial for assessing the effectiveness of the information retrieval model, with a higher $Recall@k$ indicating a greater proportion of relevant events retrieved, and a higher $Hit\ Rate@k$ suggesting that the model reliably retrieves at least one relevant event within the top $k$ results across the query events.

During stage 2 (re-ranking), the objective is to prioritize the top 10 most pertinent candidates from the 100 initially retrieved. For this purpose, we employ three metrics: Hit Rate, nDCG, and MRR. $Hit\ Rate@k$ is used again but to measure the effectiveness of surfacing at least one relevant event to the top $k$ responses when $k$ is very small. The metric is useful when considering the user experience in scenarios where only the highest-ranked results are reviewed. nDCG at $k$ ($nDCG@k$) is a rank-aware score for top-$k$ retrieved documents. It measures relevance through a rank-sensitive weight factor $\log(i+1)$, by assessing the alignment between a query's computed document ranking and the ideal ranking (Equation 4-5). For the top $k$ responses, the higher $nDCG@k \in [0,1]$ indicates that the computed ranking is closer to the ideal ranking. MRR at $k$ ($MRR@k$) reflects the ranking ability for the correct answer across $k$ queries. For each query $i$, the reciprocal rank of its responses is determined by taking the multiplicative inverse of the rank position of the first correct answer ($rank_i$), while $MRR@k$ calculates the mean of the reciprocal ranks for $k$ queries (Equation 6). A higher $MRR@k$ value, denotes a more effective ranking system, as it reflects a tendency to rank the desired relevant event closer to the top position.

## 4.3 Implementation

To establish the initial set of top 100 candidates for the GT-R model, we utilized the LlamaIndex framework designed for LLM-based applications. This enabled fast integration of OpenAI's Ada-002 embedding model to perform stage 1 retrieval. Ada was chosen due to its superior recall@100, using bi-encoding to calculate cosine similarity between event embeddings. Then went to stage 2 re-ranking. A two-step process was implemented to assess category similarity. First, NER was executed using zero-shot capabilities of the GPT-4 Turbo model. Then, cosine similarity was computed via cross-encoding to gauge the categorical alignment of events. We invoked OpenAI's API to output results directly as well-structured JSON objects. This approach streamlined the processing pipeline, significantly reducing the manual effort required for parsing and handling data, thereby easing the integration of human-in-the-loop procedures. Three models were tested for preliminary quality check including GPT-3.5-Turbo-0613, GPT-3.5-Turbo-1106, and GPT-4-1106-preview. With the same detailed CoT instructions, GPT-4 is the model that is mostly inclined to follow instructions and return desired extracted entities.

Numerical calculations for distance and latitude difference were performed using the longitude and latitude metadata associated with each event. A distance threshold was established at 500 km, with a booster factor of 2, following the LEO Network's heuristic approach. Through empirical testing with thresholds of 3, 5, and 10 degrees, an absolute latitudinal threshold of 5 degrees was set, with a corresponding latitude booster factor of 2. The determination of fusion weights for semantic similarity and category similarity was conducted through grid search. The search was configured to maintain the sum of weights at unity, with increments of 0.1 within the range from 0 to 1. The optimal weight distribution was identified with a semantic similarity weight of 0.1 and a category similarity weight of 0.9, thereby achieving the best balance for the fusion.

# 5. Result

## 5.1 Model performance

We use several zero-shot IR methods, leveraging both traditional techniques and novel LLMs, to serve as comparison with our proposed strategy for this content-based cold-start problem. Our evaluation focuses on key questions at the nexus of IR and GIS as below:
(1) Retrieval stage:

- ***RQ1:*** Input transformation: What is the optimal approach to curate input for event retrieval that captures content semantics with spatial and temporal elements?
- ***RQ2:*** LLM embedding model efficacy: Given the refined input, which models excel in the retrieval of events?

(2) Re-ranking stage:

- ***RQ3:*** GT-R efficacy: How effective is the GT-R model compared to baseline and State-of-the-art (SOTA) models in the context of spatiotemporal event re-ranking?
- ***RQ4:*** Spatial-temporal features: What effect do spatial and temporal features exert on the performance of GT-R model?

### 5.1.1 Optimizing input for event retrieval (RQ1)

To answer RQ1, we compare different input settings to examine the role of geospatial and temporal information during either sparse retrieval or dense retrieval. For sparse retrieval, we utilize BM25 (Robertson & Zaragoza, 2009), a well-established approach based on term matching and normalized document length. BM25 models serve as baselines to observe the performance improvement by exact match of added information. For dense retrieval, we use a popular LLM embedding model, "text-embedding-ada-002" from OpenAI (Neelakantan et al. 2022), which has the capability to understand context and is trained on massive datasets that potentially includes knowledge of location and date. For each input settings, we want to compare the performance of this pair of models for sparse retrieval and dense retrieval. In this way, we can understand whether a better performance of a good retrieval strategy comes from the versatility of LLM embedding models or the consideration of space and time.

Results show that the integration of spatial and temporal contents into the retrieval input significantly improves performance, as demonstrated in Table 1. When we extend the input beyond the generic semantic contents, i.e., title and summary, to include location and date, we observe a marked improvement in model performance for both sparse retrieval and dense retrieval. For example, Recall@100 of the BM25 model increases from 70.5 to 76.2 upon the addition of these dimensions. Moreover, LLM embedding models such as Ada can gain a pronounced improvement in retrieval effectiveness when inputs are structured with prefixes that denote machine-understandable contexts. For instance, optimal input formatting involves appending prefixes "Title" and "Summary" to semantic segments accordingly, and the prefix "Location" before "Kodiak, Alaska, United States", along with the prefix "Date" before "10/2/23", indicating proper spatial and temporal context. This improvement suggests a higher compatibility of structured input with the algorithms driving dense retrieval models. Using an embedding model can epitomize our pursuit of a more contextually aware and semantically rich retrieval process. The results indicate that LLM embedding models, when fed with a well-defined spatiotemporal context, are better for the retrieval task compared with models using less structured inputs.

**Table 1.** Retrieval performance (Recall@100 and Hit Rate@100) using different segment combinations as input. Best score is marked in bold.

| Company/Provider | Model | Input | Recall | Hit Rate |
|---|---|---|---|---|
| Rank-BM25 | BM25 (BM25Okapi) | Title, Summary | 70.5 | 88.2 |
| | | Title, Summary, Location | 74.2 | 89.6 |
| | | Title, Summary, Location, Date | 76.2 | 91.3 |
| OpenAI | Ada (002) | Title, Summary | 84.9 | 95.8 |
| | | Title, Summary, Location | 86.2 | 95.5 |
| | | Title, Summary, Location, Date | 86.3 | 95.9 |
| | | Title, Summary, Location, Date (with prefix) | **86.8** | **96.0** |

### 5.1.2 Retrieval model comparison (RQ2)

We employ a composite input strategy that combines title, summary, location, and date, each prefixed to enhance contextual relevance, as the basis for comparing the retrieval performance of various models. In the retrieval stage, our comparison encompasses a suite of popular embedding methods and BM 25 again as the baseline so that we can find the best retrieval model. These models include “all-MiniLM-L12-v2” from Sentence-BERT (Reimers & Gurevych, 2019), “BGE embedding” from FlagEmbedding (P. Zhang et al., 2023) and “Jina Embeddings 2” from Jina AI (Günther et al., 2023). Additionally, we engage with sophisticated commercial LLM embedding models including “embed-english-v2.0” from Cohere (Cohere, 2024a), “voyage-2” from Voyage AI (Voyage AI, 2024), and “text-embedding-ada-002” from OpenAI (Neelakantan et al., 2022), each contributing its proprietary advancement to dense retrieval.

As indicated in Table 2, dense retrieval methods demonstrate proficiency in finding a greater volume of relevant candidates with Recall@100 over 80%. This contrasts with sparse retrieval methods, which appear less effective in this context. Specifically, the BGE model achieves a Hit Rate@100 of 96.6, indicating its effectiveness in identifying at least one relevant document. Similarly, the Ada model registers the highest Recall@100 at 86.8, underscoring its superior ability to retrieve a comprehensive set of pertinent results. These results collectively suggest that dense retrieval methods, especially those using embedding techniques, have a distinct advantage in harnessing structured inputs for improved retrieval tasks. We chose the Ada model to move to the next stage.

**Table 2.** Retrieval performance (Recall@100 and Hit Rate@100) retrieving top 100 candidates of a seed event. The content segment combination of Title, Summary, Location, Date (with prefix) is used as input. Best score is marked in bold.

| Method | Company/Provider | Model | Recall | Hit Rate |
|---|---|---|---|---|
| sparse retrieval | Rank-BM25 | BM25 (BM25Okapi) | 74.7 | 90.8 |
| dense retrieval | SBERT | all-MiniLM (L12-v2) | 82.9 | 94.6 |
| | Jina AI | Jina (Embeddings 2) | 82.8 | 94.4 |
| | Voyage AI | Voyage (-2) | 86.2 | 95.4 |
| | Cohere | Cohere (embed-english-v2.0) | 86.0 | 95.8 |
| | BAAI | BGE (base-en-v1.5) | 86.4 | **96.7** |
| | OpenAI | Ada (002) | **86.8** | 96.0 |

### 5.1.3 Evaluating the GT-R model’s re-ranking efficacy (RQ3)

To understand how much of an improvement our model offers over traditional approaches that are already in use, “BM25 + distance & category booster” serves as a baseline which is currently employed by LEON as a heuristic re-ranking method. We also compare a series of models to refine our understanding of the most effective techniques for re-ranking and how various re-ranking approaches perform in the context of spatiotemporal event recommendation. For the cross-encoding approach, the comparison includes two open-source models. One model is “MS- MARCO-MiniLM" as a cross-encoding model fine-tuned on MS MARCO data set (Reimers & Gurevych, 2020), and another model is “BGE rerank” (Xiao et al., 2023) for precise result ordering. We also integrate commercial models such as “rerank-english-v2.0” from Cohere (Cohere, 2024b). For the approach directly employing LLMs as re-rankers, we test GPT-based models with different prompting strategies. For pointwise re-ranking via prompting, we test the binary relevance scoring capability of GPT3.5 Turbo based on the log probability of “Yes/No” to assess a query-candidate pair (Liang et al., 2022). For listwise re-ranking as prompting more than one candidate at a time, we run the RankGPT model (Sun et al., 2023) with a permutation generation approach and sliding window strategy to effectively re-rank a list of candidates using GPT3.5 Turbo. Together, these models provide a comprehensive spectrum of capabilities for optimizing the relevance and order of retrieved top 100 candidates to get high-quality top 10.

The performance results of comparative methods are listed in Table 3. Ada exhibited the highest recall in the retrieval phase and set a benchmark for comparison. As seen, there is a notable trend where most re-ranking models underperform after initial dense retrieval. Embedding models such as Ada are optimized for capturing the broad semantic essence of content, which results in a robust initial set of relevant articles. However, when subsequent re-ranking models, which primarily focus on semantic similarity and direct content relevance, attempt to further refine the shortlist, the marginal gains are overshadowed by the loss of spatiotemporal context. Direct use of LLMs as re-rankers struggles on this task, although considering more candidates can improve the performance. In contrast, our model GT-R surpasses all baseline and SOTA zero-shot IR techniques. It demonstrates up to 14% (47.4 vs. 41.4 as shown in Table 3) relative gains in the nDCG metric over the cutting-edge ranking model RankGPT, 6% (47.4 vs. 44.7) gains over pure dense retrieval model as strong as Ada-002, and 30% (47.4 vs. 36.4 as shown in Table 3) gains over the heuristic model used by LEO.

**Table 3.** Re-rank performance (Hit Rate@1 and 3, nDCG@10, and MRR@10) on top 100 candidates. Best score is marked in bold.

| Method | Company/ Provider | Model | Hit Rate @1 | Hit Rate @3 | nDCG @10 | MRR @10 |
|---|---|---|---|---|---|---|
| sparse retrieval followed by boosting | Rank-BM25 | BM25 (BM25Okapi) + distance & category booster | 30.2 | 53.2 | 36.4 | 43.7 |
| dense retrieval | OpenAI | Ada (002) | 39.7 | 60.0 | 44.7 | 52.2 |
| | BAAI | BGE (base-en-v1.5) | 39.2 | 59.8 | 44.6 | 51.7 |
| dense retrieval followed by re-ranking | SBERT | MS-MARCO-MiniLM (L6-v2) | 36.8 | 57.6 | 42.1 | 49.0 |
| | Cohere | Cohere rerank (english-v2.0) | 26.6 | 46.8 | 33.1 | 38.9 |
| | BAAI | BGE reranker (base) | 32.3 | 52.1 | 36.4 | 44.4 |
| | OpenAI | GPT-3.5 Turbo | 31.2 | 49.1 | 32.3 | 42.5 |
| | OpenAI | RankGPT | 36.6 | 58.2 | 41.4 | 49.0 |
| | | GT-R (our proposed method) | **42.4** | **65.3** | **47.4** | **55.8** |

### 5.1.4 Assessing the impact of semantics, space, and time to the recommendation results (RQ4)

To evaluate whether GT-R model benefits from consideration of semantics, space, and time, we conduct an ablation study by removing features of distance, latitude, time, category, and semantic respectively and report the results in Table 4. The model performance will drop when any of these re-ranking criteria is removed. The factor that causes the most significant performance decrease when removed is semantic similarity because of its critical role in the model’s ability to rank relevant events effectively. Category similarity also plays a major role but to a lesser extent than semantic similarity. Spatial and temporal features contribute to the model's performance, with geo relevance being slightly more impactful than temporal relevance, and latitudinal relevance having the least impact among the factors studied. This result indicates that considering semantics, space, and time together helps surface relevant event candidates to the top.

**Table 4.** Ablation study of GT-R model (Hit Rate@1 and 3, nDCG@10, and MRR@10).

| Feature Removed from GTR | Hit Rate @1 | Hit Rate @3 | nDCG @10 | MRR @10 | Performance Change |
|---|---|---|---|---|---|
| none | 42.2 | 65.3 | 47.4 | 55.8 | - |
| latitudinal relevance | 42.0 | 65.0 | 46.3 | 55.2 | dropped |
| temporal relevance | 41.0 | 62.8 | 47.3 | 54.4 | dropped |
| distance relevance | 39.4 | 63.6 | 45.0 | 53.2 | dropped |
| category similarity | 38.7 | 63.0 | 45.6 | 52.5 | dropped |
| semantic similarity | 33.6 | 53.9 | 39.4 | 46.1 | dropped |

## 5.2 Result examples

The performance results demonstrate the efficacy of the proposed framework in recommending contextually similar spatiotemporal events. Figure 4 visualization displays this capability, with Figure 4(a) presenting the spatial distribution of events related to a dead humpback whale spotted in Alaska. The query event is pinpointed with a red star, and the ten most similar events determined by the GT-R model are marked as brown crosses. These points are interconnected with lines to the query event, indicating their relevance. A heatmap further visualizes the concentration of recommended events. The redder the cluster, the greater the number of related events. Additionally, a 500 km radius circle around the query event and two dashed lines representing a 5-degree latitude span highlight the model's consideration of spatial proximity.

Figure 4 (b) illustrates the model's application to a landslide event in Iceland. In this case, the GT-R model successfully identifies related landslide occurrences not only within the country but also extends to recognize similar events in distant countries such as Norway and the United States. This exemplifies the model's capability in semantic search and category similarity, enabling the identification of geographically distant events underpinned by a common theme. This approach of semantics and spatiotemporal perspective underscores the GT-R model's comprehensive grasp of spatiotemporal event recommendation.

Table 5 complements Figure 4(a) by providing details of the query event involving the dead humpback whale and the top ten similar events. These events vary in location and date, enriching the understanding of the distribution of related events and their contextual linkages. For example, while some events are geographically close to the query event, others are considerably inland to the north or east such as the third recommended event of humpback whale necropsy conducted in the city of Sitka, Alaska, highlighting diverse occurrences. Notably, all recommended events occur within the 5-degree latitude relevance boosting zone, with many also falling within the 500km radius from the central event. This proximity highlights the spatial awareness of the GT-R model in aligning similar events across both spatial and temporal scales.

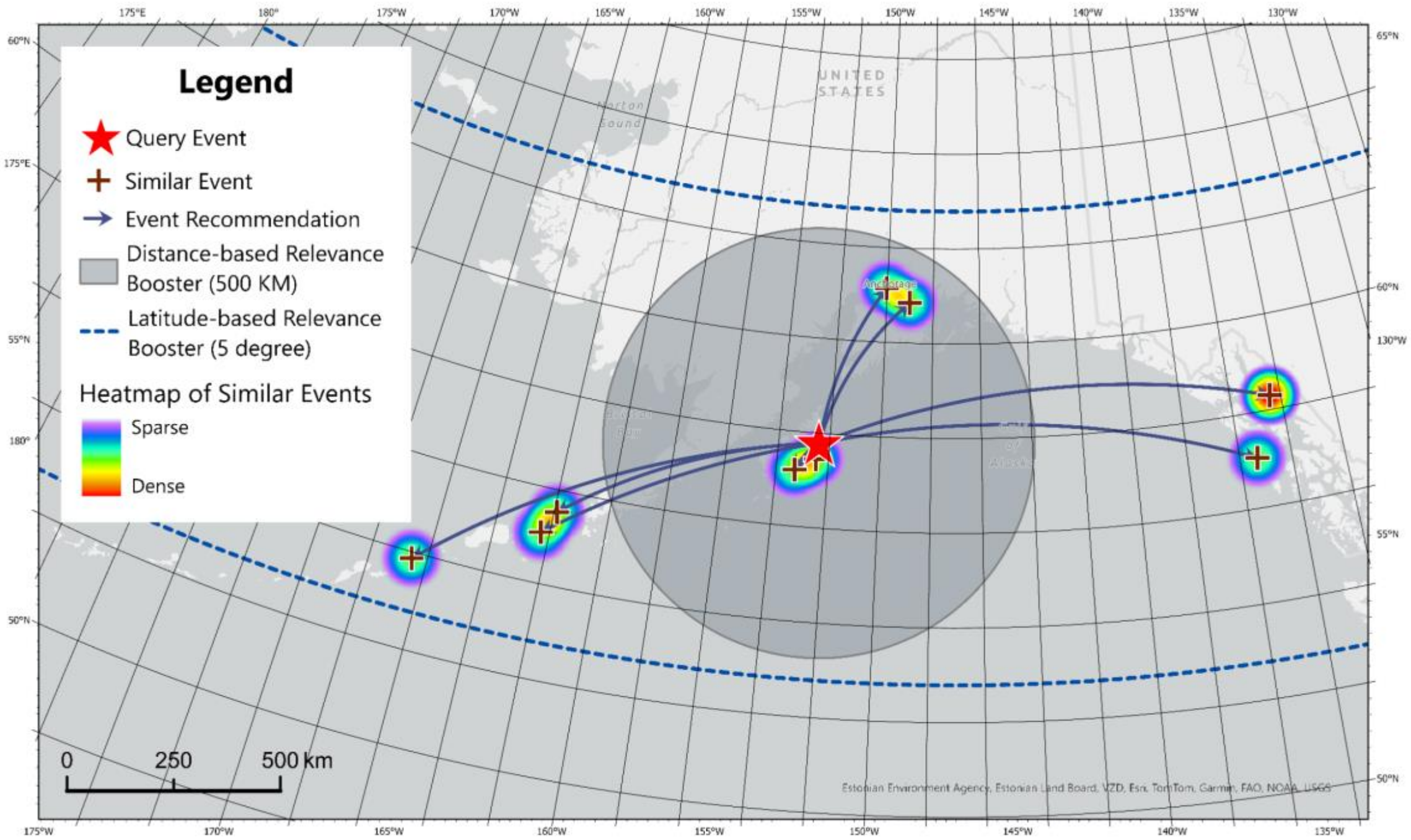

(a)

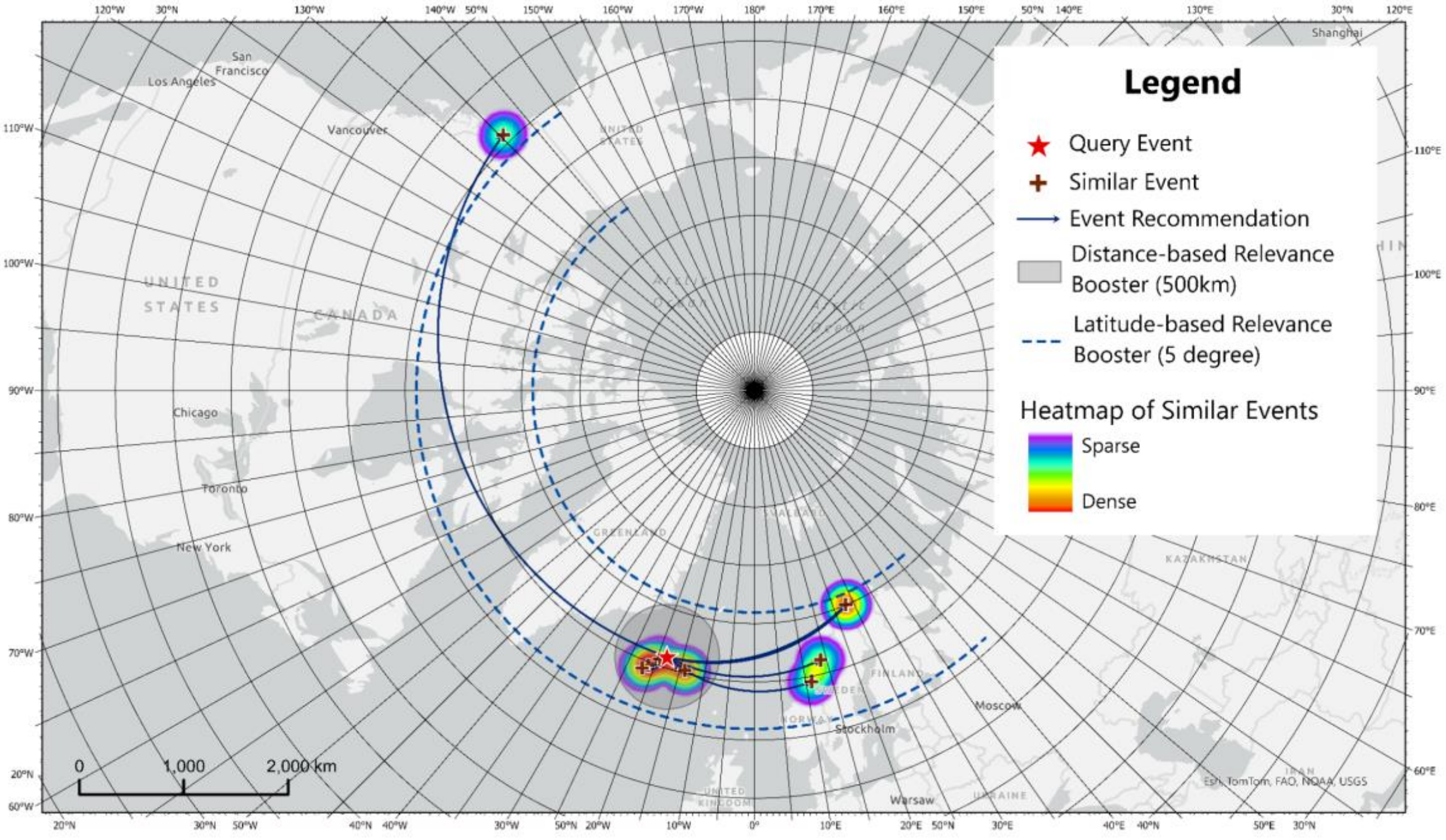

(b)

**Figure 4.** Visual representation of spatiotemporal event recommendations by GT-R Model. (a) a query event of spotting a dead whale in Alaska (red star) with its top ten similar events (brown crosses); (b) a query event regarding landslides in Finland (red start) with its top ten similar events (brown crosses). The surrounding circle represents the distance-based relevance booster (500 km). The dashed lines indicate the latitude-based relevance booster (5 degrees). The heatmap colors reflect the density of recommended events.

**Table 5.** A dead whale event in Alaska and similar events recommended by GT-R model

| | Rank | Label | Title | Location | Date |
|---|---|---|---|---|---|
| query event | | | Humpback found dead near Kodiak gets Alaska's first 2023 whale necropsy | Kodiak, Alaska, United States | 10/02/23 |
| similar events | #1 | hit | Alaska's third dead gray whale of the year reported on Kodiak coast | Kodiak, Alaska, United States | 05/23/19 |
| | #2 | | Dead Humpback Whale (Megaptera novaeangliae) | Old Harbor, Alaska, United States | 08/23/12 |
| | #3 | hit | Sitka team conducts first humpback whale necropsy in 5 years | Sitka, Alaska, United States | 03/23/21 |
| | #4 | | The dead whale floating in Cook Inlet has washed ashore at Kincaid Park | Cook Inlet, Anchorage, Alaska, United States | 09/25/17 |
| | #5 | hit | Dead Humpback Whale (Megaptera novaeangliae) | King Cove, Alaska, United States | 09/23/15 |
| | #6 | | Beluga whale found dead south of Anchorage will help scientists better understand the endangered animals | Girdwood, Alaska, United States | 05/27/21 |
| | #7 | | Dead whales wash up near Unalaska, but pandemic complicates necropsies | Unalaska, Alaska, United States | 08/27/20 |
| | #8 | | Dead humpback whale calf washes up near Juneau, may have been struck by vessel | Juneau, Alaska, United States | 08/28/23 |
| | #9 | | Dead Humpback Whale (Megaptera novaeangliae) Floating by Moller Point | King Cove Alaska | 08/18/20 |
| | #10 | | Whale's body spotted near Tenakee Inlet | Juneau, Alaska, United States | 02/10/22 |

# 6. Discussion

In this work, we have developed a framework that leverages the capabilities of LLMs combined with understanding of spatial and temporal relevance to recommend similar spatiotemporal events. The integration of LLMs with spatiotemporal insights is an innovative feature of our approach. Even when provided with basic event content without explicit geolocation and timestamps, LLM embedding models such as Ada outperform traditional sparse retrieval models, boosting recall from 70.5 to 84.9. Without the advancements in LLMs, it would be hard to achieve such a performance leap at a low cost. However, relying solely on generic analysis often means that these models lack important spatial and temporal context. As a result, embedding models can only interpret location or time information as implicitly described within the narrative. Our results during the retrieval stage suggest that incorporating specific geospatial and temporal information into the input improves model performance. Moreover, organizing inputs into machine-understandable structures, further optimizes the retrieval process. This is an intuitive yet effective strategy that aligns with the idea of prefix tuning.

Despite these advancements in retrieval, LLM-based cross-encoders and re-rankers initially show limited effectiveness in assessing environmental event similarities during the re-ranking stage, highlighting the critical role of explicitly understanding geographical and temporal nuances. Interestingly, LLMs' robust language processing capabilities enable the construction of enhanced features based on domain knowledge, which more accurately captures the semantics of events. This stage also benefits from calibrated adjustments of spatial and temporal relevance, moving beyond simple faceted search to a more sophisticated integration of these features. By merging these features, we not only achieve superior performance but also enhance the rankings beyond what is possible with pure dense retrieval methods with LLMs. In conclusion, the incorporation of LLMs along with spatial and temporal considerations during both the retrieval and re-ranking stages is vital to solving spatial problems, unlocking the power of LLMs in recommending contextually relevant events.

While showcasing on environmental events, this research demonstrates the broader applicability of our proposed method across various spatiotemporal contexts. By utilizing explicit geospatial and temporal data as integral components of the retrieval process, we demonstrate the method's adaptability to scenarios where location and date provide critical contextual information. This approach has significant potential to enhance retrieval effectiveness for events characterized merely by date and time, offering a robust framework for accurate information retrieval. Moreover, the inclusion of additional geospatial or temporal information or metadata can further enrich the retrieval process. For example, climate zones or seasonal distinctions, are often crucial for describing regions or timing weather-related events and cultural festivals. By explicitly integrating such extended context into the input, for example, by prefixing this information during the initial retrieval phase, our method can adapt and scale to accommodate richer spatiotemporal data. This prefixing strategy proves particularly effective as it minimally adjusts the frozen LLMs, guiding them toward a more contextually relevant semantic space, especially when the model has been trained to be sensitive to such prefixes. Furthermore, the strength of LLMs in tasks such as named entity recognition allows our framework to not only utilize but also to enhance entity extraction processes, moving beyond basic category tags to encompass a wider array of valuable semantic data. This capability suggests potential applications in injecting domain-specific knowledge via instructions (e.g., prompts) that could redefine the granularity and accuracy of geographic information retrieval systems implemented to fit desired scenarios. It is worth mentioning that the prompt enabling zero-shot named entity extraction is designed in a general manner, avoiding any specific references to the LEON platform or dataset. This prompting strategy mitigates potential biases from data overlap with the LLM's training data, ensuring method generalizability and results neutrality.

Despite the advancements presented by this study, the proposed framework has some limitations. One limitation is the dependency on well-defined geospatial and temporal metadata, which may not always be available or accurately captured in the datasets. This reliance on explicit contextual data can limit the applicability of our approach in scenarios where such metadata is sparse or non-existent. Additionally, while the LLMs employed show significant capabilities in processing and interpreting complex datasets, their performance can still be influenced by the inherent biases and the quality of data used to train the model originally. These models may not always generalize well across different contexts or domains without substantial tuning or adaptation. Looking forward, extending the capabilities of our framework could involve enhancing model transparency and adaptability. Conducting stress tests will be essential to identify and mitigate biases by exposing the model to diverse and challenging scenarios within the domain. Furthermore, leveraging few-shot learning will enable us to fine-tune the LLMs with minimal but highly relevant data, enhancing their domain adaptability and reducing potential biases. These steps are beneficial to improving the robustness of such system, ensuring it performs reliably across varied contexts and contributes to advancing the transparency and effectiveness of LLM applications in geographic information retrieval.

Another point worth noting is about how to further improve time relevance. Our model occasionally returns a candidate list with low recall, particularly in scenarios involving a series of events. For example, in the case of a query event titled “Wildfire smoke affecting millions of Canadians expected to linger for at least 2 days,” the recall scores at top 10 and top 100 are both 0, indicating a failure to retrieve relevant events during the initial stage. This shortfall is partly attributed to the model’s inability to recognize and prioritize events linked by a significant phenomenon. In this case, it was the widespread impact of Canada’s extreme 2023 fire season. Despite that top 100 candidates are indeed fire or wildfire events primarily focused on air quality, the cascading events of that megafire (Linley et al., 2022) started in Canada get overshadowed. A possible solution worth exploring involves introducing an expert model to handle the seasonality of significant events such as megafires, enabling a mixture strategy of time relevance assessment for dynamic adjustment. Furthermore, there is room for enhancing the GT-R model’s performance through methodological advancements. Our current approach leverages LLMs to enrich category tags with expert-level insight, suggesting the potential for future research to automate knowledge generation, such as automatic tag creation. However, the hallucination issue remains a concern, where LLMs produce off-topic or fictitious content. Adopting a domain-specific controlled vocabulary could mitigate this risk, ensuring the generation of accurate and relevant topic tags.

Besides, there’s still room to improve how spatial context is considered. Techniques such as retrieve-augment-generation or agentic API calling techniques are promising for enriching spatial context by providing information on land use or surrounding Points of Interest (POIs), and by considering differences such as coastal versus inland, and rural versus urban. This would add a layer of relevance to the proposed framework. To effectively incorporate this context factor, it is important to define clear metrics to assess the impact of coast-to-land differences on the relevance of environmental events. Additionally, utilizing auxiliary datasets or knowledge bases of land cover and land change can help identify coastal and inland regions. Exploring advanced LLM techniques that can incorporate and reflect complex geospatial distinctions in event retrieval and reranking would further enhance the framework’s effectiveness.

# 7. Conclusion

As news becomes increasingly important for sharing information and raising awareness about environmental events related to climate change, effectively recommending similar events on news platforms has become crucial for readers seeking to explore related content. In our study, we developed a retrieval and re-ranking framework that leverages the capabilities of LLMs to enhance the recommendation of spatiotemporal event news. We explained the methodological details of our approach, the Geo-Time Re-ranking (GT-R) model, and systematically compared it with existing techniques, including both sparse and dense retrieval models, and re-ranking models. Our findings show that the proposed model achieves significant improvements with up to 40% gains over the heuristic solution, i.e., keyword matching with distance boosting recommendation method currently adopted in the LEO system. Its performance also surpasses several state-of-the-art re-ranking models. The GT-R model's superior performance is primarily attributed to two key factors: (1) it explicitly considers multiple relevance ranking criteria, including spatial proximity, temporal association, semantic similarity, and category-instructed similarity, and (2) the strategic use of LLM models to construct and enrich semantic features. By incorporating LLMs into the GT-R framework, we leverage their advance natural language processing capabilities to interpret and contextualize the complexity of content semantics and spatiotemporal information. Thus, this approach allows the GT-R model to recognize and prioritize events that are not only semantically similar but also contextually relevant in terms of their occurrence in space and time. The significant and broader impact of our work lies in its potential to assist climate change experts for automatically annotating relevant news articles or events by employing cutting-edge AI techniques. Through the integration of LLMs and a geo-time perspective, we envision it can not only facilitate the rapid identification of similar spatiotemporal events but also to accelerate the synthesis of indicators assessing the health and vulnerability of ecosystems.

# Acknowledgement

This research is supported in part by Google.org's Impact Challenge for Climate Innovation Program, OpenAI's Researcher Access Program, as well as the U.S. National Science Foundation (Grants No. 2230034, 1853864, and 2120943).